\documentstyle[12pt]{article}
\def\singlespace {\smallskipamount=3.75pt plus1pt minus1pt
                  \medskipamount=7.5pt plus2pt minus2pt
                  \bigskipamount=15pt plus4pt minus4pt
                  \normalbaselineskip=15pt plus0pt minus0pt
                  \normallineskip=1pt
                  \normallineskiplimit=0pt
                  \jot=3.75pt
                  {\def\smallskip {\vskip\smallskipamount}}
                  {\def\medskip   {\vskip\medskipamount}}
                  {\def\bigskip   {\vskip\bigskipamount}}
                  {\setbox\strutbox=\hbox{\vrule
                    height10.5pt depth4.5pt width 0pt}}
                  \parskip 7.5pt
                  \normalbaselines}
\def\middlespace {\smallskipamount=5.825pt plus1.5pt minus1.5pt
                  \medskipamount=11.25pt plus3pt minus3pt
                  \bigskipamount=22.5pt plus6pt minus6pt
                  \normalbaselineskip=22.5pt plus0pt minus0pt
                  \normallineskip=1pt
                  \normallineskiplimit=0pt
                  \jot=5.825pt
                  {\def\smallskip {\vskip\smallskipamount}}
                  {\def\medskip   {\vskip\medskipamount}}
                  {\def\bigskip   {\vskip\bigskipamount}}
                  {\setbox\strutbox=\hbox{\vrule
                    height15.75pt depth6.75pt width 0pt}}
                  \parskip 7.25pt
                  \normalbaselines}
\def\dblspc {\smallskipamount=7.5pt plus2pt minus2pt
                  \medskipamount=15pt plus4pt minus4pt
                  \bigskipamount=30pt plus8pt minus8pt
                  \normalbaselineskip=30pt plus0pt minus0pt
                  \normallineskip=2pt
                  \normallineskiplimit=0pt
                  \jot=7.5pt
                  {\def\smallskip {\vskip\smallskipamount}}
                  {\def\medskip   {\vskip\medskipamount}}
                  {\def\bigskip   {\vskip\bigskipamount}}
                  {\setbox\strutbox=\hbox{\vrule
                    height21.0pt depth9.0pt width 0pt}}
                  \parskip 15.0pt
                  \normalbaselines}

\def\be{\begin{equation}}
\def\lan{\left\langle}
\def\ran{\right\rangle}
\def\ee{\end{equation}}

\def\barr{\begin{array}}
\def\earr{\end{array}}
\def\cn9{\nonumber\\[15pt]}
\def\nn8{\nonumber\\[10pt]}
\def\l{\left}
\def\r{\right}

\def\dis{\displaystyle}
\def\ed{\end{document}}
\def\btl{\tilde{b}}
\def\bdag{b^\dagger}
\def\fsp{\frac{1}{2}}

\begin{document}
\middlespace
\begin{flushleft}
{\bf O(12) limit and complete classification of symmetry schemes in 
proton-neutron interacting boson model}
\vskip 0.75cm
V.K.B. Kota  \\
Physical Research Laboratory, Ahmedabad \,\,380 009,
India 
\end{flushleft}
\vskip 1cm

\noindent {\bf Abstract:} It is shown that the proton-neutron interacting
boson model (pnIBM) admits new  symmetry limits with $O(12)$ algebra which
break $F$-spin but preserves the $F_z$ quantum number $M_F$.  The generators
of O(12) are derived and the quantum number $v$ of $O(12)$ for a given boson
number $N$ is determined by identifying the corresponding quasi-spin
algebra. The $O(12)$ algebra generates two symmetry schemes and for both  of
them, complete classification of the basis states and typical spectra  are
given. With the $O(12)$ algebra identified, complete classification of pnIBM
symmetry limits with good $M_F$ is established. 


\vskip 0.5cm


\newpage
\begin{flushleft}
{\bf 1. Introduction}
\end{flushleft}
\noindent

The most significant aspect of the interacting boson model (IBM) of
even-even nuclei \cite{Ia-87} is the dynamical symmetries of the model. In
its most elementary version with scalar ($s$) and quadrupole ($d$) bosons,
the model (called IBM-1) admits the well established (they are now part of
several text books \cite{Ta-00}) vibrational $U(5)$, rotational $SU(3)$  and
$\gamma$-unstable $O(6)$ symmetries starting with the $U(6)$ spectrum
generating algebra (SGA) (the 6 in $U(6)$ corresponds to the sum of one
degree of freedom from $s$ bosons (with angular momentum $\ell=0$) and five
from $d$ bosons ($\ell=2$). Its extended version, with proton ($\pi$) and
neutron ($\nu$) degrees of freedom attached to $s$ and $d$ bosons, the
proton-neutron interacting boson model (pnIBM or IBM-2) admits dynamical
symmetries starting from its $U(12)$ SGA \cite{Ia-87,Va-85,Fr-94}. An
important development here is the introduction of the so-called $F$-spin;
$\pi$ and $\nu$ bosons are considered as two projections of a spin-$\fsp$
boson. Going beyond IBM-1 and IBM-2 models, in the last five years, the
dynamical symmetries of the isospin invariant IBM-3 \cite{Ko-98} and
spin-isospin invariant IBM-4 \cite{Ko-00,I4-00} models are also being
studied as they are shown to have applications for nuclei near the proton
drip line. Although the IBM-1 model was introduced nearly 20 years back,
remarkably there is now new interest in re-examining the symmetries of these
models with developments and interest in quantum chaos and phase
transitions. In the context of the former, the importance of
$\overline{O(6)}$ and $\overline{SU(3)}$ algebras \cite{ Ku-99} is brought
out and for the later the so-called $E(5)$ and $X(5)$ symmetries are
introduced \cite{Ia-01}. Also new interpretations for the $SU(3)$ and $O(6)$
limits are proposed with higher-order interactions \cite{Va-01}. Then an
immediate question that arises is whether there are new symmetries in pnIBM
that are not recognized before. This question is answered in the affirmative
in this paper by identifying and analyzing the $O(12)$ limit group
chains of pnIBM.

The $sd$ boson pnIBM with $\pi-\nu$ degrees of freedom is a standard model
for analyzing the properties of heavy even-even nuclei with protons and
neutrons occupying different oscillator shells \cite{Ia-87,Ot-00}. The SGA
of pnIBM is $U(12)$ as a single boson carries 12 degrees of
freedom (6 from $s$ and $d$ and two from $\pi$ and $\nu$) in this model.
Then there are the well known $U(6) \otimes SU_F(2)$ with the $SU_F(2)$
algebra generating $F$-spin and the $U_{\pi}(6) \oplus U_{\nu}(6)$ symmetry
limits in this model; in addition there are also the $U_s(2) \oplus U_d(10)$
symmetry limits (see Section 4). The various group chains starting  from
$U(6) \otimes SU_F(2)$ and $U_\pi(6) \oplus U_\nu(6)$ are identified and
studied in great detail in the past; see for example
\cite{Ia-87,Va-85,Fr-94}. Let us point out that IBM-1 model corresponds to
the $F=N/2$ states in pnIBM where $N$ is the total number of bosons.  The
$F=N/2-1$ states are the so-called mixed symmetry states. In rotational
$SU(3)$ nuclei they correspond to the now well known scissors states that
are seen in many nuclei \cite{Sc-85}. It should be emphasized that there are
many new experiments, with the advent of the EUROBALL cluster detector, in
the last five years in identifying the mixed symmetry states of pnIBM in
$O(6)$ type nuclei, for example in $^{196}$Pt, $^{134,136}$Ba and $^{94}$Mo
isotopes \cite{Vo-00}. Though the focus is in identifying good $F$-spin
states in $O(6)$ nuclei, it is well known that $F$-spin is broken in many
situations \cite{Ta-89}. In this paper, following the $O(18)$ and $O(36)$
symmetry limits of IBM-3 and IBM-4 models \cite{Ko-98,Ko-00}, it is
identified that pnIBM admits a $O(12)$ limit with broken $F$-spin but good
$F_z$ quantum number $M_F$. It should be mentioned that, although the
existence of $O(12)$ limit is mentioned in the past in \cite{Ko-99,La-97},
in this paper for the first time the $O(12)$ symmetry chains,  as they are
closely related to $O(6)$ nuclei, are analyzed in any detail. Section 2
gives the generators and the quadratic Casimir operator of $O(12)$ by
identifying the corresponding quasi-spin algebra; also discussed here is the
closely related $O(10)$ algebra in $d$ boson space. Two group chains are
possible with $O(12)$ and Section 3 gives classification of states and
typical spectra for both of them. In Section 4 complete classification of
pnIBM symmetry limits with good $M_F$ is discussed in detail. Finally,
Section 5 gives concluding remarks.

\begin{flushleft}
{\bf 2. $O(12)$ symmetry in pnIBM}
\end{flushleft}

\begin{flushleft}
2.1 {\it Preliminaries}
\end{flushleft}

\noindent The pnIBM, with proton - neutron degrees of freedom, can be
described in general in terms of $\pi-\nu$ representation or the equivalent
$F$-spin representation with the identification $\l.\l| \pi \r.\ran = \l.\l|
F=\fsp\;,\;M_F=\fsp \r.\ran=\l.\l| \fsp\;\fsp \r.\ran$ and $\l.\l| \nu
\r.\ran = \l.\l| F=\fsp\;,\;M_F=-\fsp \r.\ran = \l.\l| \fsp\;-\fsp\r.\ran$
\cite{Ia-87}. In the $F$-spin representation, given the one boson creation
and destruction operators $b^{\dag}_{\ell, m_\ell;\;\fsp, m_f}$ and
$\btl_{\ell, m_\ell;\;\fsp, m_f} = (-1)^{ \ell+ m_\ell + \fsp + m_f}\;
b_{\ell, -m_\ell;\;\fsp, -m_f}$, the 144 double tensors
$\l(\bdag_{\ell,\fsp} \btl_{\ell^\prime,\fsp} \r)^{L_0, F_0}_{M_0, M_{F_0}}$
generate the $U(12)$ SGA; note that for us $\ell=0(s)$ or $2(d)$. Similarly,
in the $\pi-\nu$ representation, given the one boson creation and
destruction operators $b^{\dag}_{\ell, m_\ell;\;\rho}$ and $\btl_{\ell,
m_\ell;\;\rho} = (-1)^{ \ell+ m_\ell}\; b_{\ell, -m_\ell;\;\rho}$;
$\rho=\pi$ or $\nu$, the 144 operators $\l(\bdag_{\ell,
\rho}\btl_{\ell^\prime, \rho^\prime}\r)^{L_0}_{M_0}$ generate the $U(12)$
algebra. These results follow directly from the following commutation
relations,
\be
\barr{l}
\l[ \l(\bdag_{\ell_1,\fsp} \btl_{\ell_2,\fsp} \r)^{L_{12}, F_{12}}_{
M_{12}, M_{F_{12}}}
\;\;\;\l(\bdag_{\ell_3,\fsp} \btl_{\ell_4,\fsp} \r)^{L_{34}, F_{34}}_{M_{
34}, M_{F_{34}}} \r]_-\;\;= \;\;\cn9
\dis\sqrt{(2L_{12}+1)(2L_{34}+1)(2F_{12}+1)(2F_{34}+1)}\;(-1)^{1+\ell_1+
\ell_4}
\dis\sum_{L_0\;,\;F_0}\;\;\cn9
\lan L_{12} M_{12} \;\;L_{34} M_{34}\;\mid L_0 M_0 \ran
\lan F_{12} M_{F_{12}} \;\;F_{34} M_{F_{34}}\;\mid F_0 M_{F_0} \ran \;\;
(-1)^{L_0+F_0} \;\;\times \cn9
\l[ \l\{\barr{ccc} L_{12} & L_{34} & L_0 \\ \ell_4 & \ell_1 & \ell_2 
\earr \r\}
\l\{\barr{ccc} F_{12} & F_{34} & F_0 \\ \fsp & \fsp & \fsp \earr \r\}\;
\l(\bdag_{\ell_1,\fsp} \btl_{\ell_4,\fsp} \r)^{L_{0}, F_{0}}_{
M_{0}, M_{F_{0}}}\;\;\delta_{\ell_2\,\ell_3} \r. \cn9
- (-1)^{\ell_1+\ell_2+\ell_3+\ell_4+L_{12}+L_{34}+L_0+F_{12}+F_{34}+F_0}\;
\times\; \cn9
\l. \l\{\barr{ccc} L_{12} & L_{34} & L_0 \\ \ell_3 & \ell_2 & \ell_1 
\earr \r\}
\l\{\barr{ccc} F_{12} & F_{34} & F_0 \\ \fsp & \fsp & \fsp \earr \r\}\;
\l(\bdag_{\ell_3,\fsp} \btl_{\ell_2,\fsp} \r)^{L_{0}, F_{0}}_{
M_{0}, M_{F_{0}}}\;\;\delta_{\ell_1\,\ell_4} \r] 
\earr
\ee
\noindent and
\be
\barr{l}
\l[ \l(\bdag_{\ell_1,\rho_1} \btl_{\ell_2,\rho_2} \r)^{L_{12}}_{M_{12}}
\;\;\;\l(\bdag_{\ell_3,\rho_3} \btl_{\ell_4,\rho_4} \r)^{L_{34}}_{M_{34}} 
\r]_-\;\;= \;\;\cn9
\dis\sqrt{(2L_{12}+1)(2L_{34}+1)}\;(-1)^{\ell_1+\ell_4}
\dis\sum_{L_0}\;\lan L_{12} M_{12} \;\;L_{34} M_{34}\;\mid L_0 M_0 \ran
\;(-1)^{L_0} \;\;\times \cn9
\l[ \l\{\barr{ccc} L_{12} & L_{34} & L_0 \\ \ell_4 & \ell_1 & \ell_2 
\earr \r\}
\;\l(\bdag_{\ell_1,\rho_1} \btl_{\ell_4,\rho_4} \r)^{L_{0}}_{M_{0}}\;\;
\delta_{\ell_2\,\ell_3} \; \delta_{\rho_2 , \rho_3} \r. \cn9
- (-1)^{\ell_1+\ell_2+\ell_3+\ell_4+L_{12}+L_{34}+L_0}\;
\l. \l\{\barr{ccc} L_{12} & L_{34} & L_0 \\ \ell_3 & \ell_2 & \ell_1 
\earr \r\}
\l(\bdag_{\ell_3,\rho_3} \btl_{\ell_2,\rho_2} \r)^{L_{0}}_{M_{0}}\;\;
\delta_{\ell_1\,\ell_4} \; \delta_{\rho_1 , \rho_4}\r] 
\earr
\ee
Dynamical symmetry limits of pnIBM correspond to the group chains starting
with $U(12)$ generating  $N$ and ending with $O_L(3) \otimes \l[SU_F(2)
\supset O_{M_F}(2)\r]$ or $O_L(3) \otimes O_{M_F}(2)$ generating states with
good $(N,L,F,M_F)$ or only $(N,L,M_F)$ respectively; note that $N=(N_\pi +
N_\nu)$ and $M_F=(N_\pi - N_\nu)/2$; $N_\pi$ and $N_\nu$ are proton and
neutron boson numbers respectively. Before going further it is useful to
write down the number, $F$-spin and angular momentum ($L$) 
operators\footnote{in general it is easily seen that,
$\l(\bdag_{\ell_1,\fsp} \btl_{\ell_2,\fsp}\r)^{L_0,0}_{M_0,0}= \frac{1}{
\sqrt{2}}\;\sum_{\rho=\pi,\nu}\;\l(\bdag_{\ell_1,\rho}
\btl_{\ell_2,\rho}\r)^{L_0}_{M_0}$},
\be
\barr{rcl}
\hat{n} & = & {\hat{n}}_s + {\hat{n}}_d = {\hat{n}}_\pi + {\hat{n}}_\nu \nn8
& = & \dis\sum_{\ell=0,2}\;\dis\sqrt{2(2\ell+1)}\;
\l(\bdag_{\ell,\fsp} \btl_{\ell,\fsp}\r)^{0,0}_{0,0} \;
= \; \dis\sum_{\rho=\pi,\nu}\;
\l(s^\dagger_\rho {\tilde{s}}_\rho\r)
+ \dis\sqrt{5} \dis\sum_{\rho=\pi,\nu}\;
\l(d^\dagger_\rho {\tilde{d}}_\rho\r)^0 \nn8
& = & \l[s^\dagger_\pi {\tilde{s}}_\pi
+ \dis\sqrt{5} \l(d^\dagger_\pi {\tilde{d}}_\pi\r)^0 \r] +
\l[s^\dagger_\nu {\tilde{s}}_\nu
+ \dis\sqrt{5} \l(d^\dagger_\nu {\tilde{d}}_\nu\r)^0 \r]
 \nn8
\\
L^1_\mu & = & \dis\sqrt{20} \l(d^\dagger_\fsp {\tilde{d}}_\fsp \r)^{
1,0}_{\mu,0}
= \dis\sqrt{10} \sum_{\rho=\pi,\nu}\;\l(d^\dagger_\rho {\tilde{d}}_\rho
\r)^{1}_{\mu} \nn8
\\
F^1_\mu & = & \dis\frac{1}{\dis\sqrt{2}} \dis\sum_{\ell=0,2}\;\dis\sqrt{
(2\ell+1)}\;\l(\bdag_{\ell,\fsp} \btl_{\ell,\fsp}\r)^{0,1}_{0,\mu}\;; \nn8
F^1_0 & = & \dis\frac{1}{2} \l(\l[{\hat{n}}_{s;\pi}+{\hat{n}}_{d;\pi}\r] -
\l[{\hat{n}}_{s;\nu} + {\hat{n}}_{d;\nu}\r]\r)\;, \nn8
F^1_1 & = & -\dis\frac{1}{\dis\sqrt{2}} \l(\l[s^\dagger_\pi s_\nu\r] +
\dis\sum_m \l[d^\dagger_{m;\pi} d_{m;\nu}\r]\r)\,,\;\;
F^1_{-1} \; = \; \dis\frac{1}{\dis\sqrt{2}} \l(\l[s^\dagger_\nu s_\pi\r] +
\dis\sum_m \l[d^\dagger_{m;\nu} d_{m;\pi}\r]\r)\;.
\earr
\ee
The proton $s$ and $d$ boson and neutron $s$ and $d$ boson number operators
${\hat{n}}_{s;\pi}$,  ${\hat{n}}_{s;\nu}$,  ${\hat{n}}_{d;\pi}$ and
${\hat{n}}_{d;\nu}$ are defined by the third equality in (3) and
similarly, the decompositions of $L$ into $\pi$ and $\nu$ parts and $F$
components into $s$ and $d$ parts follow immediately from (3).

At the primary level, as pointed out in the introduction, identified by the
first sub-algebra of $U(12)$, pnIBM has four symmetry limits \cite{Ko-99}:
(i) $U(6) \otimes SU_F(2)$; (ii) $U_{\pi}(6) \oplus U_{\nu}(6)$; (iii)
$U_s(2) \oplus U_d(10)$; (iv) $O(12)$.  With the condition that $N$, $L$ and
$M_F=(N_\pi-N_\nu)/2$ must be good quantum numbers, there will be no other
chains in pnIBM except those related to (i)-(iv).  Complete classification
of group chains with good $(N,L,M_F)$ in pnIBM will be discussed in detail
in Section 4. In the present section the $O(12)$ algebra is studied in
detail. As the $O(12)$ algebra is defined in  $sd$ boson
space, it is more appropriate to start first with the corresponding $O(10)$
algebra in $d$ boson space.

\begin{flushleft}
2.2 {\it $O(10)$ algebra in $d$ boson space}
\end{flushleft}

In $d$ boson space the SGA is $U(10)$ and starting with it there are two
chains: (i) $U(10) \supset \l[U(5) \supset O(5) \supset O_L(3)\r] \otimes
\l[SU_F(2) \supset O_{M_F}(2)\r]$ where $F$-spin is good; (ii) $U(10)
\supset O(10) \supset \l[O(5) \supset O_L(3)\r] \otimes O_{M_F}(2)$ where
only $M_F$ is good. Here we are concerned with (ii), the $O(10)$ 
chain; chain (i) is considered in section 4 ahead. It is known
that $U(M)$ admits $O(M)$ as a sub-algebra, thus $U(10) \supset O(10)$ is
always possible. But the question is whether there is a $O(10)$ that
preserves $L$ and $M_F$. The answer is in the affirmative and this is seen
from the generators of $O(10)$ which are identified to be,
\be
\barr{rcl}
O(10) & : & A^{L=1,3}_\mu=\l(d^\dagger_\pi \tilde{d}_\nu\r)^{1,3}_\mu\,,\;\;
B^{L=1,3}_\mu=\l(d^\dagger_\nu \tilde{d}_\pi\r)^{1,3}_\mu\,,\nn8
& & C^{L=0-4}_\mu=\l[\l(d^\dagger_\pi \tilde{d}_\pi\r)^L_\mu + 
(-1)^{1+L}\,\l(d^\dagger_\nu \tilde{d}_\nu\r)^L_\mu\r]
\earr
\ee
It is seen from (2) that $\l[A^{L_1}_{\mu_1}\; A^{L_2}_{\mu_2}\r]_- =0$,
$\l[B^{L_1}_{\mu_1} \; B^{L_2}_{\mu_2}\r]_- =0$, $\l[A^{L_1}_{\mu_1}
\; B^{L_2}_{\mu_2}\r]_-$ is a sum of $C^L$'s, $\l[A^{L_1}_{\mu_1}
\; C^{L_2}_{\mu_2}\r]_-$ is a sum of $A^L$'s, $\l[B^{L_1}_{\mu_1}
\; C^{L_2}_{\mu_2}\r]_-$ is a sum of $B^L$'s and finally $\l[C^{L_1}_{\mu_1}
C^{L_2}_{\mu_2}\r]_-$ is a sum of $C^L$'s. The $C^1_\mu$ generate $\vec{L}$,
$C^0$ generates $M_F$ and $C^{L=1,3}_\mu$ generate $O(5)$ in the chain
$O(10) \supset \l[O(5) \supset O_L(3)\r] \otimes O_{M_F}(2)$. It is clear
that, as $F^1_{d;\pm}$ operators are not in (4), the $O(10)$ chain
breaks $F$-spin. In order to understand the physical meaning of $O(10)$ and
determine the $O(10)$ irreducible representations (irreps) contained in the
symmetric irreps $\{N_d\}$ of $U(10)$ ($N_d$ is number of $d$ bosons),
following \cite{Ko-96}, the corresponding quasi-spin $SU_{S;d}(2)$
algebra is constructed. The generators $S_\pm(d)$ and $S_0(d)$
of $SU_{S;d}(2)$ and their commutation relations are,
\be
\barr{l}
S_+(d) = \dis\sqrt{5} \l(d^\dagger_\pi d^\dagger_\nu\r)^0\,,
\;\;S_-(d) = \dis\sqrt{5}
\l(\tilde{d}_\pi \tilde{d}_\nu\r)^0\,,\;\;S_0(d)=\frac{\l(5+
{\hat{n}}_d\r)}{2} \nn8
\l[S_+(d)\;S_-(d)\r]_-=-2S_0(d)\;,\;\;\; \l[S_0(d)\;S_\pm(d)\r]_- =
\pm S_\pm(d)
\earr
\ee
With $\{S(d)\}^2=S_0(d)(S_0(d)-1)-S_+(d)S_-(d)$ and $S_0(d)$ defining
$\l.\l|S_d M_{S_d}\,\alpha^\prime \r.\ran$ basis ($\alpha^\prime$ 
labels states with the
same $S_d$ and $M_{S_d}$ values in the $d$ boson space) and the results
$M_{S_d}=\frac{1}{2} (5+N_d)$, $S_d=M_{S_d}, M_{S_d}+1, \ldots$ and $\lan
\{S(d)\}^2 \ran^{S_d,M_{S_d}}=S_d(S_d-1)$ give, using $S_d=\frac{1}{2}(
5+v_d)$,
\be
\barr{c}
S_+(d)S_-(d) = 5 \l(d^\dagger_\pi d^\dagger_\nu\r)^0 \l(\tilde{d}_\pi
\tilde{d}_\nu\r)^0 = \dis\sum_{L_0} (-1)^{L_0} \,
\l(d^\dagger_\pi \tilde{d}_\pi\r)^{L_0} \cdot \l(d^\dagger_\nu
\tilde{d}_\nu\r)^{L_0}\;, \nn8
\lan S_+(d)S_-(d) \ran^{S_d \,M_{S_d}} =
\lan S_+(d)S_-(d) \ran^{N_d \,v_d} = \frac{1}{4} \l(N_d-v_d\r) \l(N_d+
v_d+8\r)\;;\nn8
v_d=N_d, N_d-2, \ldots,\;0\;\;\mbox{or}\;\;1
\earr
\ee
The relationship between $SU_{S;d}(2)$ and $O(10)$ is derived by examining
the quadratic Casimir operators of $U(10)$ and $O(10)$,
\be
\barr{rcl}
C_2(U(10)) & = &
\dis\sum_k \l(d^\dagger_\pi \tilde{d}_\pi\r)^{k} \cdot \l(d^\dagger_\pi
\tilde{d}_\pi\r)^{k}\;+\;
\dis\sum_k \l(d^\dagger_\nu \tilde{d}_\nu\r)^{k} \cdot \l(d^\dagger_\nu
\tilde{d}_\nu\r)^{k} \nn8
& & + \dis\sum_k \l(d^\dagger_\pi \tilde{d}_\nu\r)^{k} 
\cdot \l(d^\dagger_\nu \tilde{d}_\pi\r)^{k}\;+\;
\dis\sum_k \l(d^\dagger_\nu \tilde{d}_\pi\r)^{k} \cdot \l(d^\dagger_\pi
\tilde{d}_\nu\r)^{k} \nn8
C_2(O(10)) & = &
2\dis\sum_{k=1,3} \l(d^\dagger_\pi \tilde{d}_\nu\r)^{k} 
\cdot \l(d^\dagger_\nu \tilde{d}_\pi\r)^{k}\;+\;
2\dis\sum_{k=1,3} \l(d^\dagger_\nu \tilde{d}_\pi\r)^{k} 
\cdot \l(d^\dagger_\pi\tilde{d}_\nu\r)^{k} \nn8
& & + \dis\sum_L\;\l[ \l(d^\dagger_\pi \tilde{d}_\pi\r)^L + (-1)^{1+L}
\l(d^\dagger_\nu \tilde{d}_\nu\r)^L \r] \cdot 
\l[ \l(d^\dagger_\pi \tilde{d}_\pi\r)^L + (-1)^{1+L}
\l(d^\dagger_\nu \tilde{d}_\nu\r)^L \r] 
\earr
\ee
Following \cite{Ko-96} it can be recognized that the four terms in
$C_2(U(10))$ give $\l[{\hat{n}}_{d;\pi} ({\hat{n}}_{d;\pi}-1) + 5
{\hat{n}}_{d;\pi}\r]$, $\l[{\hat{n}}_{d;\nu} ({\hat{n}}_{d;\nu}-1) + 5
{\hat{n}}_{d;\nu}\r]$, $\l[{\hat{n}}_{d;\pi}{\hat{n}}_{d;\nu} + 5
{\hat{n}}_{d;\pi}\r]$ and $\l[{\hat{n}}_{d;\pi}{\hat{n}}_{d;\nu} + 5
{\hat{n}}_{d;\nu}\r]$ respectively. Similarly  $2\dis\sum_{k=1,3}
\l(d^\dagger_\pi \tilde{d}_\nu\r)^{k} \cdot \l(d^\dagger_\nu
\tilde{d}_\pi\r)^{k} =  {\hat{n}}_{d;\pi}{\hat{n}}_{d;\nu} + 4
{\hat{n}}_{d;\pi} - S_+(d)S_-(d)$ gives
\be
C_2(O(10))=-4S_+(d)S_-(d) + {\hat{n}}_d({\hat{n}}_d+8)
\ee
Now applying (6) gives finally,
\be
\barr{c}
\l.\l| \barr{ccc} U(10) & \supset & O(10) \\ N_d & & v_d \earr \r.
\ran\;,\;\;
v_d=N_d, N_d-2, \ldots,\;0\;\;\mbox{or}\;\;1 \nn8
\\
\lan C_2(U(10)) \ran^{N_d\, v_d} = N_d(N_d+9)\;,\;\;
\lan C_2(O(10)) \ran^{N_d\, v_d} = v_d(v_d+8)
\earr
\ee
Thus the pairs in the $O(10)$ limit are $\pi-\nu$ boson pairs and
\be
\l.\l| N_d\;v_d\;\alpha^\prime\r.\ran = \l\{\dis\frac{(v_d+4)!}{\l[(N_d-
v_d)/2\r]!\;
\l[(N_d+v_d+8)/2\r]!}\r\}^{1/2} \l[\dis\sqrt{5} \l(d^\dagger_\pi 
d^\dagger_\nu\r)^0\r]^{(N_d-v_d)/2}\;\l.\l| v_d\;v_d\;
\alpha^\prime\r.\ran
\ee

\begin{flushleft}
2.3 {\it O(12) generators and the corresponding quasi-spin algebra}
\end{flushleft}

In $sd$ boson space, following the results in Section 2.2, it is natural to
expect the appearance of $U(12) \supset O(12)$ algebra. From Section 2.2 it
is clear that the 45 generators $A^{L=1,3}_\mu$, $B^{L=1,3}_\mu$ and
$C^{L=0-4}_\mu$ of $O(10)$ in $d$ boson space (see (4)) and the generator
$D^0=(s^\dagger_\pi \tilde{s}_\pi - s^\dagger_\nu \tilde{s}_\nu)
=2F^1_{s;0}$
of $O(2)$ in $s$ boson space will be in the $O(12)$ algebra. Then the
remaining 20 generators of $O(12)$ need to be identified. From the
generators of $O(6)$ in $U(6)$ of IBM-1, it is easily seen that
$E^2_\mu=\l[\l(s^\dagger_\pi \tilde{d}_\nu\r) +  \alpha \l(d^\dagger_\pi
\tilde{s}_\nu\r)\r]^2_\mu$ and $F^2_\mu=\l[\l(s^\dagger_\nu \tilde{d}_\pi\r)
+  \beta \l(d^\dagger_\nu \tilde{s}_\pi\r)\r]^2_\mu$ will be in the $O(12)$
algebra. The commutators $\l[A^L_\mu \;F^2_{\mu^\prime}\r]_-$ $\l[B^L_\mu
\;E^2_{\mu^\prime}\r]_-$ immediately give the remaining 10 generators
$G^2_\mu=\l[\l(s^\dagger_\nu \tilde{d}_\nu\r) +  \gamma \l(d^\dagger_\pi
\tilde{s}_\pi\r)\r]^2_\mu$ and $H^2_\mu=\l[\l(s^\dagger_\pi \tilde{d}_\pi\r)
+  \delta \l(d^\dagger_\nu \tilde{s}_\nu\r)\r]^2_\mu$. By evaluating all the
commutators, using (2), between the 66 generators $A^{L=1,3}_\mu$,
$B^{L=1,3}_\mu$, $C^{L=0-4}_\mu$, $D^0$, $E^2_\mu$, $F^2_\mu$, $G^2_\mu$ and
$H^2_\mu$ it is seen for example that $[A\;F]_-$ gives $G$, $[A\;H]_-$ gives
$E$ and $[E\;F]_-$ gives a sum of $D^0$ and $C^L$ only if
$\alpha=\beta=\gamma=\delta$ and $\alpha^2=1$. Applying these conditions it
is seen that the following 66 operators generate the $O(12)$ algebra in
pnIBM,
\be
\barr{rcl}
O(12) & : & A^{L=1,3}_\mu=\l(d^\dagger_\pi \tilde{d}_\nu\r)^{1,3}_\mu\,,\;\;
B^{L=1,3}_\mu=\l(d^\dagger_\nu \tilde{d}_\pi\r)^{1,3}_\mu\,,\nn8
& & C^{L=0-4}_\mu=\l[\l(d^\dagger_\pi \tilde{d}_\pi\r)^L_\mu + 
(-1)^{1+L}\,\l(d^\dagger_\nu \tilde{d}_\nu\r)^L_\mu\r]\,, \nn8
& & D^0=(s^\dagger_\pi \tilde{s}_\pi - s^\dagger_\nu \tilde{s}_\nu)\,, \nn8
& & E^2_\mu=\l[\l(s^\dagger_\pi \tilde{d}_\nu\r) + 
\alpha \l(d^\dagger_\pi \tilde{s}_\nu\r)\r]^2_\mu \;,\;\; 
F^2_\mu=\l[\l(s^\dagger_\nu \tilde{d}_\pi\r) + 
\alpha \l(d^\dagger_\nu \tilde{s}_\pi\r)\r]^2_\mu \,,\nn8
& & G^2_\mu=\l[\l(s^\dagger_\nu \tilde{d}_\nu\r) + 
\alpha \l(d^\dagger_\pi \tilde{s}_\pi\r)\r]^2_\mu\;,\;\;
H^2_\mu=\l[\l(s^\dagger_\pi \tilde{d}_\pi\r) + 
\alpha \l(d^\dagger_\nu \tilde{s}_\nu\r)\r]^2_\mu \,,\nn8
& & \alpha = \pm 1 \,.
\earr
\ee
Thus there are two $O(12)$ algebras, one with $\alpha=1$ and other with
$\alpha =-1$. Now we will construct the corresponding quasi-spin algebras.

Combining the $SU_{S:d}(2)$ quasi-spin algebra in $d$-space and 
the corresponding algebra $SU_{S;s}(2)$ in $s$-space defined by 
\be
S_+(s)=s^\dagger_\pi s^\dagger_\nu\;,\;\; S_-(s)=\tilde{s}_\pi 
\tilde{s}_\nu\;,\;\;S_0(s)=\frac{(1+\hat{n}_s)}{2}\;,
\ee
it is straightforward to introduce the quasi-spin $SU_S(2)$ algebra in the 
total $sd$-space,
\be
S_+ = S_+(s) + \beta\, S_+(d)\;,\;\; S_- = S_-(s) + \beta\, S_-(d)\;,\;\;
S_0 =  \frac{(6+\hat{n})}{2}\;;\;\;\; \beta=\pm 1
\ee
The relationship between $\alpha$ in (11) and $\beta$ in (12) is established
ahead. With the quasi-spin algebra (13), we have $\l.\l|N\,v\,\alpha^\prime
\r.\ran$ states
exactly as in (6,9) and $\lan S_+S_-\ran^{N,v} = (N-v)(N+v+10)/4$.
In order to see this, let us first define $C_2(O(12)$,
\be
C_2(O(12)) = C_2(O(2)) + C_2(O(10)) + \alpha\l[E \cdot F + F \cdot E +
G \cdot H + H \cdot G\r]
\ee
where $C_2(O(2))$ is 
\be
C_2(O(2)) = D^0 D^0 = n_s^2 - 4S_+(s)S_-(s)\,,
\ee
$C_2(O(10))$ is defined by (8) and $\alpha$ is defined in (11). Recognizing
that
\be
\barr{c}
E \cdot F + F \cdot E = 2\l[S_+(s) S_-(d) + S_+(d) S_-(s)\r] + \alpha \l[
5n_s + n_d + 2 n_{s;\pi} n_{d;\nu} + 2 n_{d;\pi} n_{s;\nu} \r]\,, \nn8
G \cdot H + H \cdot G = 2\l[S_+(s) S_-(d) + S_+(d) S_-(s)\r] + \alpha \l[
5n_s + n_d + 2 n_{s;\nu} n_{d;\nu} + 2 n_{d;\pi} n_{s;\pi} \r]
\earr
\ee
and using (8), (13) and (15) it is sen that $C_2(O(12))$ can be written in
terms of $S_+S_-$ only when $\alpha=-\beta$. Then finally, with $\alpha=-\beta$,
\be
C_2(O(12)) = -4 S_+ S_- + {\hat{n}}(\hat{n}+10)
\ee
Thus the $O(12)$ defined by the generators in (11) correspond to the 
quasi-spin algebra defined by (13) when $\alpha=-\beta$. With this we have, 
just as in (9,10),
\be
\barr{l}
\l.\l| \barr{ccc} U(12) & \supset & O(12) \\ N & & v \earr \r.
\ran\;,\;\;
v=N, N-2, \ldots,\;0\;\;\mbox{or}\;\;1 \nn8
\\
\lan C_2(O(12)) \ran^{N,\,v} = v(v+10) \nn8
\l.\l| N\;v\;\alpha^\prime\r.\ran = \l\{\dis\frac{(v+5)!}{\l[(N-v)/2\r]!\;
\l[(N+v+10)/2\r]!}\r\}^{1/2} \l[s^\dagger_\pi s^\dagger_\nu + \dis\sqrt{5}
\beta \l(d^\dagger_\pi
d^\dagger_\nu\r)^0\r]^{(N-v)/2}\;\l.\l| v\;v\;\alpha^\prime\r.\ran
\earr
\ee
where $\beta=\pm 1$ and the $\alpha$ in the $O(12)$ generators in (11) is
related to $\beta$ by $\alpha=-\beta$.

\begin{flushleft}
{\bf 3. Spectra in $O(12) \supset O(6) \otimes O(2)$ and $O(12)
\supset O(2) \oplus O(10)$ limits}
\end{flushleft}

The $O(12)$ algebra admits $O(6) \otimes O(2)$ and $O(2) \oplus O(10)$
subalgebras with good $M_F$. In both cases one can write down the complete
group chains with good $(N,L,M_F)$. Hereafter these two chains are called  
$O(12) \supset O(6) \otimes O(2)$ and $O(12) \supset O(2) \oplus O(10)$
limits respectively of pnIBM. Let us point out that, in addition to $M_F$, 
the $O(12) \supset O(2) \oplus O(10)$ limit also preserves $M_{F_s}$ and
$M_{F_d}$ and hence it is more restrictive.

\begin{flushleft}
{\it 3.1 $O(12) \supset O(6) \otimes O(2)$ limit}
\end{flushleft}

The group chain and irrep labels in the $O(12) \supset O(6) \otimes O(2)$
limit are given by,
\be
\l.\l|\barr{ccccccccccccc} U(12) & \supset & O(12) & \supset & [ & O(6) &
\supset & O(5) & \supset & O_L(3) & ] &  \otimes & O_{M_F}(2) \\
\{N\} & & \l[v\r] & & & \l[\sigma_1 \sigma_2\r] & & \l[v_1 v_2\r]  & & L
& & & M_F \earr\r.\ran
\ee
The $O(6)$ algebra is generated by the 15 generators $C^{L_0=1,3}_\mu$ and
$G^2_\mu + \beta^\prime H^2_\mu$ and similarly the $O_{M_F}(2)$ is generated
by $D^0+\sqrt{5} C^0$ where $C^{L_0}$, $D^0$, $G^2$ and $H^2$ are defined in
(11); the $O(5)$ and $O_L(3)$ algebras are generated by $C^{L_0=1,3}_\mu$
and $C^1_\mu$ respectively. For the $O(6)$ algebra in (19) to be same as the
$O(6)$ in the $U(6) \otimes SU_F(2) \supset O(6) \otimes O_{M_F}(2)$ limit
of pnIBM (as stated earlier, this limit was studied in detail in the past
\cite{Ia-87,Va-85}), one needs the conditions $\alpha=1$ in (11) and
$\beta^\prime=1$ in $G^2_\mu + \beta^\prime H^2_\mu$ generators. With these
conditions met, it is possible to compare the results in the these two
limits and derive (see ahead) the new structures implied by (19). Before the
results for irrep labels are given, it should be pointed out that for a
given nucleus $N$, $L$ and $M_F$ are always good quantum numbers. The $N
\rightarrow v$ reduction problem was already solved in Section 2 (see Eq.
(18)) and the $v
\rightarrow \l[\sigma_1 \sigma_2\r]\,M_F$ reductions are given in 
Appendix A; note that here Table 1 with $r=6$ will apply. 
The rule for $\l[\sigma_1 \sigma_2\r]
\rightarrow \l[v_1 v_2\r]$ is well known \cite{Fr-94,Wy-70,Ko-97a},
$\sigma_1 \geq v_1 \geq \sigma_2 \geq v_2 \geq 0$. Finally  $\l[v_1 v_2\r]
\rightarrow L$ can be solved using (A4) and the general solution for
$\l[\tau\r]_{O(5)} \rightarrow L$. For example  $\l[0\r]_{O(5)} \rightarrow
L=0$, $\l[1\r]_{O(5)} \rightarrow L=2$, $\l[2\r]_{O(5)} \rightarrow L=2,4$,
$\l[11\r]_{O(5)} \rightarrow L=1,3$, $\l[3\r]_{O(5)} \rightarrow L=0,3,4,6$
and $\l[21\r]_{O(5)} \rightarrow L= 1,2,3,4,5$. Using these irrep reductions
and writing the hamiltonian as a linear combination of the quadratic
Casimir operators of the groups in (19) one can construct the typical
spectrum in the $O(12) \supset O(6) \otimes O(2)$ limit. The hamiltonian and
the energy formula in this limit are,
\be
\barr{l}
H \; = \; E_0(N,M_F) + a_1 C_2(O(12)) + a_2 C_2(O(6)) + a_3 C_2(O(5)) 
+ a_4 C_2(O(3)) \nn8
E\l(N,v,\l[\sigma_1 \sigma_2\r],\l[v_1 v_2\r],L,M_F\r) =  
E_0(N,M_F) +
a_1\,v(v+10) + a_2\,\l[\sigma_1(\sigma_1+4) + \sigma_2(\sigma_2+2)\r] \nn8
\;\;\;\;\;\;\;\;\;\;\;\;\;\;\;\;\;\;\;\;\;\;\;\;\;\;\;\;\;\;\;\;\;
\;\;\;\;\; + a_3\,\l[v_1(v_1+3) + 
v_2(v_2+1)\r] + a_4\,L(L+1)
\earr
\ee
The operator form for $C_2(O(12))$ and the formula for its eigenvalues are
given in Section 2. The corresponding results for $C_2(O(6))$, $C_2(O(5))$
and $C_2(O(3))$ are easy to write down \cite{Ia-87,Fr-94,Ko-97a}. In order
to get IBM-1 like states to be lowest, for a given $v$ we need $\l[\sigma_1
\sigma_2\r]=\l[v,0\r]$ to be lowest and therefore $a_2 < 0$ in (20). In
order to get the ground $L=0,2,4 \ldots$ band correctly we need $a_3 > 0$
and $a_4 \sim 0$. With these restrictions it is seen that the condition $a_1
>0$ gives a spectrum similar to the spectrum in the $U(6) \otimes SU_F(2)
\supset O(6) \otimes O_{M_F}(2)$ limit. As an example, for $N=6$ and
$M_F=-1$ (then $N_\pi=2$, $N_\nu=4$) the typical spectrum in the $O(12)
\supset O(6) \otimes O(2)$ limit is shown in Fig. 1 and this should be
compared with the $U(6) \otimes SU_F(2) \supset O(6) \otimes O_{M_F}(2)$ limit
spectrum given in Fig. 4 of Ref. \cite{Va-85}. Firstly the states with  the
$O(6)$ irreps $[6]$ and $[51]$ in Fig. 1 belong to $v=6$ (i.e. $v=N$) and 
therefore it is not
possible in general to separate them too far. Due to this, as seen from 
Fig. 1, the $[51]$ states start appearing around 1.5 MeV excitation.
Typically states with the irrep $[6]$ are IBM-1 states and the $[51]$ states
are the mixed symmetry states.  In the $U(6) \otimes SU_F(2) \supset O(6)
\otimes O_{M_F}(2)$ limit the $[6]$ and $[51]$ $O(6)$ irreps belong to
different $U(6)$ irreps and therefore in this limit it is possible to split
them far by using the $U(6)$ Casimir operator (the Majorana operator
\cite{Ia-87,Va-85}). With this in the $U(6) \otimes SU_F(2) \supset O(6) 
\otimes O_{M_F}(2)$ limit the mixed symmetry states are expected
around 3 MeV excitation as found in many nuclei. Unlike this, in the
$O(12) \supset O(6) \otimes O(2)$ limit they are expected to appear around
1.5 to 2 MeV as in Fig. 1. Therefore to find empherical examples for this
symmetry limit one has to look for
$O(6)$ type even-even nuclei with $1^+$ states (see Fig. 1) appearing around
1.5 MeV.  In fact there are many such nuclei \cite{So-92} and in order to
establish their structure one need to study their $B(E2)$'s.  It should be
added that the $\l[\sigma_1 \sigma_2\r]=[N]$ and $[N-1,1]$ states with $v=N$
in the $O(12) \supset O(6) \otimes O(2)$ limit will have same structure as
in the $U(6) \otimes SU_F(2) \supset O(6) \otimes O_{M_F}(2)$  limit as the
corresponding $U(6)$ irreps are uniquely determined. Additional signatures
for the $O(12) \supset O(6) \otimes O(2)$ limit come from the $v=N-2$ (in
Fig. 1 they correspond to $v=4$) states which should start appearing around
2.2 - 2.5 MeV excitation (around this, states with $v=N$ and $O(6)$ irrep
$[N-2,2]$ also will start appearing but they are not shown in the figure)
and they will have one $sd$ boson $\pi-\nu$ pair (see (18)). Therefore these
states carry much more definite signatures of $O(12)$. 
For further understanding of
the $O(12) \supset O(6) \otimes O(2)$ limit, it is necessary to study
the structure of the eigenstates in this limit in terms
of the amount of $F$-spin mixing
they contain and also derive formulas for $B(E2)$'s and $B(M1)$'s between 
the states in this limit. They will be addressed in a future publication.

\begin{flushleft}
{\it 3.2 $O(12) \supset O(2) \oplus O(10)$ limit}
\end{flushleft}

The group chain and irrep labels in the
$O(12) \supset O(2) \oplus O(10)$ limit are given by,
\be
\barr{l}
\l|\barr{ccccccccccccc} U(12) & \supset & O(12) & \supset & [ & 
O(10) & \supset & \{ & O(5) & \supset & O_L(3) & \} &  \otimes \nn8
\{N\} & & \l[v\r] & & & \l[v_d\r] & & & \l[v_1 v_2\r] & & L \earr \r.
\nn8
\\
\l.\barr{cccccccc}
O_{M_{F_d}}(2) & ] & \oplus & O_{M_{F_s}}(2) & \supset & O_L(3) & 
\otimes & O_{M_F}(2) \nn8
M_{F_d} & & & M_{F_s} & & L & & M_F=M_{F_s}+M_{F_d} \earr\ran
\earr
\ee
The generators of all the groups in (21) will follow from the results in
Sections 2 and 3.1. The irrep labels in (21) are determined as follows.
It is seen from (18) that $v=N,N-2,\ldots,0\;\mbox{or}\;1$.
The $v \rightarrow (v_d,M_{F_s})$ is given by the rule
\cite{Ko-97} $v=v_s+v_d+2k,\;
k=0,1,2,\ldots$ where $v_s=2 \l|M_{F_s}\r|$. The rules for
$v_d \rightarrow \l[v_1 v_2\r]\,M_{F_d}$ are given in 
Appendix A; note that here Table 1 with $r=5$ will apply. The
$\l[v_1 v_2\r] \rightarrow L$ reductions are same as in the case of
$O(12) \supset O(6) \otimes O(2)$ limit. Now let us consider the hamiltonian
and the energy formula in the  $O(12) \supset O(2) \oplus O(10)$ limit,
\be
\barr{l}
H=E_0(N,M_F)+b_1 C_2(O(12)) +b_2 C_2(O(10)) +b_3 C_2(O(5))  \nn8
+b_4 C_2(O(3))+ b_5 \l(F^1_{d;0}\r)^2 \nn8
\\
E\l(N,v, v_d, \l[v_1 v_2\r],L, M_{F_d}, M_F\r)=
E_0(N,M_F) + b_1 v(v+10) + b_2 v_d(v_d+8) \nn8
+ b_3 \l[v_1(v_1+3) + 
v_2(v_2+1)\r] + b_4 L(L+1) + b_5 (M_{F_d})^2
\earr
\ee
Assuming $b_1 < 0$, the $v=N$ states will be lowest in energy. Then choosing
$b_2>0$ in principle a phonon spectrum can be obtained. For $v_d=0$ one has $L=0$ with
$M_{F_d}=0$. Note that $M_{F_s}=\pm\frac{(v-v_d)}{2}, \pm\frac{(v-v_d-2)}{2},
\ldots, 0\;\mbox{or}\;\pm\fsp$ and given $M_{F_d}$ the $M_{F_s}$ value
must be $M_{F_s}=M_F-M_{F_d}$. For $v_d=1$ (one phonon excitation)
one has $\l[v_1 v_2\r]=[1]$,
$L=2$ with $M_{F_d}=\pm \fsp$. Similarly for $v_d=2$ one has $L=0,2,4$ with
$M_{F_d}=\pm 1$ and $L=(2,4), (1,3)$ with $M_{F_d}=0$. 
These results follow from
Table 1 as $\l[v_1 v_2\r]=[0] \rightarrow L=0$, $[1] \rightarrow L=2$,
$[2] \rightarrow L=2,4$ and $[11] \rightarrow L=1,3$. This construction
extends to $v_d=3$ etc. Then there will be similar states with $v=N-2$ at
energies higher than those of $v=N$ states and so on. With these,
by choosing $b_5=0$ one has, a ground $0^+$, 1-phonon excited $2^+$ with
$M_{F_d}=\pm\fsp$, 2-phonon $(0^+,2^+,4^+)$ with $M_{F_d}=\pm 1$ and 
$(2^+,4^+)$, $(1^+,3^+)$ with $M_{F_d}=0$. Thus, because of the
degeneracies due to good $(M_{F_s},M_{F_d})$ in this symmetry limit, the
spectrum is unrealistic. However by adding a term 
$b_6\l[\fsp v_d-M_{F_d}\r]\l[v_d+R_1 M_{F_d}+R_2\r]$ with $b_6$, $R_1$ and 
$R_2$ being some constants, in the energy formula (22), it is possible
to lift the $M_{F_d}$ degeneracies to give a spectrum that looks realistic.
At present a two-body interaction with eigenvalues 
$\l[\fsp v_d-M_{F_d}\r]\l[v_d+R_1 M_{F_d}+R_2\r]$ could not be
constructed. In conclusion, it is probable that the  $O(12) \supset O(2) 
\oplus O(10)$  limit will not be seen in real nuclei.

\begin{flushleft}
{\bf 4 Complete classification of pnIBM symmetry limits with good $M_F$}
\end{flushleft}

With the $O(12)$ algebra identified and studied in Sections 2 and 3, it is
natural to address the question of complete classification of the symmetry
schemes (group-subgroup chains) in pnIBM. As already pointed out, they are
associated with the four $U(12)$ subalgebras (i) $U(6) \otimes SU_F(2)$,
(ii) $U_{\pi}(6) \oplus U_{\nu}(6)$, (iii) $U_s(2) \oplus U_d(10)$ and (iv)
$O(12)$. In the $U(6) \otimes SU_F(2)$ limit, the $U(6)$ algebra is
generated by $\l(\bdag_{\ell_1,\fsp} \btl_{\ell_2, \fsp}\r)^{L_0, 0}_{M_0,
0}$; $\ell_1, \ell_2=0, 2$ and $SU_F(2)$ by the $F$-spin operators $F^1_\mu$
in (3). All the group chains in this limit are well known 
\cite{Ia-87,Va-85} and they correspond to the  the sub-algebras $G$'s in
$U(12) \supset \l[U(6) \supset G \supset \cdots \supset O_L(3) \r] \otimes
\l[SU_F(2) \supset O_{M_F}(2)\r]$; $G=U(5),SU(3)$, $O(6)$. Obviously all
these chains preserve $(N,L,F,M_F)$ (note that we are not showing $O_L(3)
\supset O_{M_L}(2)$ as $L$ is an exact symmetry). In the $U_\pi(6) \oplus
U_\nu(6)$ limit the $U_\rho(6)$ generators follow easily from (2) and they
are  $\l(\bdag_{\ell_1,\rho} \btl_{\ell_2, \rho}\r)^{L_0}_{M_0}$;
$\ell_1, \ell_2=0, 2$ and $\rho=\pi,\nu$. The boson numbers $N_\rho$ are
generated by $U_\rho(6)$ and therefore the group chains in the $U_\pi(6) \oplus
U_\nu(6)$ limit will always preserve $M_F$. The various group chains in
this limit are obtained by writing down all the $U_\rho(6)$ subalgebras
with good $L_\rho$ ($\rho=\pi,\nu)$, then coupling the $\pi-\nu$
algebras at some level and further reducing this coupled algebra to
$O_L(3)$. All these group chains are well known \cite{Ia-87} and they are of
the form $U(12) \supset \l[U_\pi(6) \supset \ldots G_\pi \supset \ldots
\r] \oplus \l[U_\nu(6) \supset \ldots G_\nu \supset \ldots \r] \supset 
G_{\pi+\nu} \ldots \supset O_L(3)$. In summary, the  
$U(6) \otimes SU_F(2)$ symmetry
limit group chains preserve $(N,L,F,M_F)$ and the $U_\pi(6) \oplus U_\nu(6)$ 
symmetry limit group chains preserve only  $(N,L,M_F)$ and all these group 
chains are known before.\footnote{In the $U_\pi(6) \oplus U_\nu(6)$
limit, the special case of coupling at the $U_\rho(6)$ level itself, i.e.
coupling the $U_\pi(6)$ and $U_\nu(6)$ to give $U_{\pi+\nu}(6)$, is
equivalent to $U(6) \otimes SU_F(2)$. This special case is often used in
literature to describe the $U(6) \otimes SU_F(2)$ symmetry schemes
\cite{Ia-87,Va-85}.} 

In the $U_s(2) \oplus U_d(10)$ limit, the $sd$ boson space is decomposed into
$s$ and $d$ spaces so that not only $N$ but both $N_s$ (generated by
$U_s(2)$) and $N_d$ (generated by $U_d(10)$) are good quantum numbers. The
$U_s(2)$ generates $s$-boson $F$-spin $F_s=N_s/2$. The  $U_d(10)$ admits
two subalgebras as pointed out in Section 2.2 and with this there are
two group chains in the $U_s(2) \oplus U_d(10)$ limit,
\be
\barr{l}
\l|\barr{cccccccccccccc} U(12) & \supset & 
U_s(2) & \oplus & [ & U_d(10) & \supset & \{ & U(5) & 
\supset & O(5) & \supset & O_L(3) & \} \nn8
\{N\} & & \{N_s\};F_s=N_s/2 & & & \{N_d\} & & & \{f_1,f_2\} & & 
\l[v_1 v_2\r] & & L \earr \r.
\nn8
\\
\l. \barr{ccccccc} \otimes & SU_{F_d}(2) & ] & \supset & SU_F(2) & \supset 
& O_{M_F}(2) \nn8
& F_d=(f_1-f_2)/2  & & & \vec{F}=\vec{F_s}+\vec{F_d} & & M_F \earr \ran
\earr
\ee
\vskip 0.1cm
\be
\barr{l}
\l|\barr{ccccccccccccc} U(12) & \supset &  [ & U_s(2) & \supset & 
O(2)_{M_{F_s}} & ] & \oplus & 
[ & U_d(10) & \supset & O(10) & \supset \nn8
\{N\} & & & \{N_s\},F_s=N_s/2 & & M_{F_s} & & & & \{N_d\} & & [v_d] 
\earr \r.
\nn8
\\
\l.\barr{cccccccccc} \{ & O(5) & \supset & O_L(3) & \} & \otimes & 
O_{M_{F_d}}(2) & ] & \supset & O_{M_F}(2) \nn8
& \l[v_1 v_2\r] & & L & & & M_{F_d} & & & M_F=M_{F_s}+M_{F_d} \earr \ran
\earr
\ee
In the first chain (23) $F$-spin is good and by examining the irrep reductions
(basis states) it is easily seen that it is same as the
$U(6) \otimes SU(2)$ limit with $U(6) \supset U(5)$ (see \cite{Va-85} and 
Fig. 2 in this reference); note that in (23), $N=N_s+N_d$, $f_1+f_2=N_d$, 
$f_1 \geq f_2 \geq 0$. The second chain (24)
(hereafter called $U(2) \oplus [U(10) \supset O(10)]$ limit) is a new 
group chain in pnIBM. 
For this chain, the irrep reductions $N_d \rightarrow v_d$ and $v_d \rightarrow
\l[v_1 v_2\r] M_{F_d}$ follow from the results in Section 2 and Appendix A;
results in Table 1 with $r=5$ will apply here. The $\l[v_1 v_2\r] \rightarrow
L$ reductions are given in Section 3.1. It is straight forward to write down
the hamiltonian and energy formula in this limit. Just as in the case of
$O(12) \supset O(2) \oplus O(10)$, it is easily seen that the spectrum in the
present case also will be unrealisistic (with degeneracies due to good 
$M_{F_d}$). Thus both 
$U(2) \oplus [U(10) \supset O(10)]$ and $O(12) \supset O(2) \oplus O(10)$
which preserve $(M_{F_s}$, $M_{F_d})$, may not be seen in real nuclei but they
should be useful for chaos and phase transition studies (see ahead).

In summary, combining the symmetry schemes in the $U(6) \otimes SU_F(2)$,
$U_{\pi}(6) \oplus U_{\nu}(6)$ and $U_s(2) \oplus U_d(10)$ limits with the
two $O(12)$ symmetry limits analyzed in Section 3, one has the complete
classification of symmetry schemes with good $(N,L,M_F)$ in pnIBM. 
 
\begin{flushleft}
{\bf 5. Conclusions}
\end{flushleft}

Proton-neutron interacting boson model admits a new $O(12)$ symmetry limit
which breaks $F$-spin but preserves the $F_z$ quantum number $M_F$.  The
$O(12)$ algebra is analyzed in detail, for the first time in this paper, by
identifying the corresponding quasi-spin algebra. With $O(12)$ there are two
symmetry limits in pnIBM, $O(12) \supset O(6) \otimes
O(2)$ and $O(12) \supset O(2) \oplus O(10)$ limits. In both cases complete
classification of the basis states and typical energy spectra are given. It
is argued that some  $O(6)$ type ($\gamma$ soft) nuclei may exhibit 
the  $O(12) \supset O(6) \otimes O(2)$ limit and two important signatures 
here are the appearance of $O(6)$ $\l[N-1,1\r]$ states around 1.5 MeV
excitation and the $0_3^+$ (or $0_4^+$) states around 2.5 MeV  with a
correlated $\pi-\nu$ boson pair. Search for empirical examples is under
progress.

Searching for complete classification of pnIBM symmetry schemes, it is found
that, within the $U_s(2) \oplus U_d(10)$ algebra of $U(12)$, there is a new
$U(2) \oplus  [U(10) \supset O(10)]$ limit. This may be relevant for $U(5)$
(vibrational) type nuclei. For the three new symmetry 
limits discussed in this
paper,  $O(12) \supset O(6) \otimes O(2)$, $O(12) \supset O(2) \oplus O(10)$
and $U(2) \oplus [U(10) \supset O(10)]$ given by (19), (21) and (24)
respectively (note that they all preserve $M_F$ and in general break the
$F$-spin), results for electromagnetic transition strengths ($B(E2)$'s and
$B(M1)$'s) and structure of wavefunctions in terms of the amount of $F$-spin
mixing they contain, will be presented elsewhere. The group theoretical 
problems needed for these are being solved.

With the $O(12)$ limit studied in Section 3, another important problem 
addressed and solved in this paper is the complete
classification of pnIBM symmetry schemes with good $M_F$.  Let us point out
that a major application of the complete classification is in the studies of
quantum chaos and phase transitions in finite quantum systems where one can
use pnIBM as a model. Such studies with great success are  carried out using
IBM-1 \cite{Ku-99,Ia-01,Al-90,Cj-01} and only  recently a beginning is
made in this direction using pnIBM \cite{Ca-00}.

\begin{flushleft}
{\bf Acknowledgments}
\end{flushleft}

Collaboration with Mr Gautum Kumar for his help in generating some of the
results in this paper is acknowledged.

\begin{flushleft}
{\bf Appendix A}
\end{flushleft}

The problem of $\l[\tau\r] \rightarrow \l[\tau_1\,\tau_2\r]\,M_F$ irrep 
reductions in the group-subgroup chain
$$
\l.\l| \barr{ccccccc} U(2r) & \supset & O(2r) & \supset & O(r) & \otimes &
O(2) \\
\{N\} & & \l[\tau\r] & & \l[\tau_1\,\tau_2\r] & & M_F \earr \r.\ran\;;
\;\;\;\tau = N, N-2, N-4, \dots, 0\;\mbox{or}\;1\;\;\;\;\;\;\;\;\;(A1)
$$
is solved by using the group chain
$$
\barr{l}
\l.\l| \barr{ccccccccc} U(2r) & \supset & U(r) & \otimes & SU(2) &
\supset & O(r) & \otimes & O(2) \\
\{N\} & & \{f_1 f_2\} & & F & & \l[\tau_1\,\tau_2\r] & &
M_F \earr \r.\ran\;; \nn8
f_1+f_2=N,\;\; f_1 \geq f_2 \geq 0,\;\; F=(f_1-f_2)/2,\;\; M_F=-F, 
(-F+1), \ldots, F\;\;\;\;\;\;\;\;\;\;\;\;\;(A2)
\earr
$$
The $\{f_1\,f_2\} \rightarrow \l[\tau_1\,\tau_2\r]$ reduction in (A2) is
obtained by the well known rules for the $U(r)$ and $O(r)$ Kronecker
($\otimes$) products \cite{Wy-70} and the $U(r) \supset O(r)$ reductions for
the symmetric $U(r)$ irreps,
$$
\l\{f_1\r\}_{U(r)} \otimes \l\{f_2\r\}_{U(r)} - \l\{f_1+1\r\}_{U(r)}
\otimes \l\{f_2-1\r\}_{U(r)} =
\l\{f_1, f_2\r\}_{U(r)} \;\;\;\;\;\;\;\;\;\;\;\;\;\;\;\;\;\;\;(A3)
$$
$$
\l[\kappa\r]_{O(r)} \otimes \l[\ell\r]_{O(r)} = \dis\sum_{p=0}^{\ell}
\dis\sum_{q=0}^{\ell-p}\;\l[\kappa-\ell+p+2q, p\r]_{O(r)} \oplus \;\;,\;
\ell \leq \kappa\;\;\; \;\;\;\;\;\;\;\;\;\;\;\;\;\;\;\;\;\;\;\;\;(A4)
$$
$$
\l\{f\r\}_{U(r)} \longrightarrow \l[f\r]_{O(r)} \oplus \l[f-2\r]_{O(r)} \oplus
\ldots \oplus \l[0\r]_{O(r)}\;\;\mbox{or}\;\;\l[1\r]_{O(r)}
\;\;\;\;\;\;\;\;\;\;\;\;\;\;\;\;\;\;\;\;\;\;\;(A5)
$$
By writing all allowed $\{f_1 f_2\}$ in (A2) for a given $N$ and then
applying (A3), (A5) and (A4) in that order will give $\{N\} \rightarrow
\l[\tau_1\,\tau_2\r]\,M_F$ reductions. Starting with $N=1,3,5 \ldots$ and by
successive substraction of the  $\{N\} \rightarrow  \l[\tau_1\,\tau_2\r]\,
M_F$ reductions will give, via $N \rightarrow \tau$ in (A1), the
$\l[\tau\r]_{O(2r)} \rightarrow \l[\tau_1\,\tau_2\r]_{O(r)}\,
\l(M_F\r)_{O(2)}$ irrep reductions for odd $N$. Similarly, starting with
$N=0,2,4,\ldots$ will give the reductions for even $N$. This procedure is
easily implemented on a computer. Table 1 gives the results for $\tau \leq
6$. From the table it is seen that, in general,
$$
\barr{rcl}
\l[\tau\r]_{O(2r)} & \longrightarrow \l[\tau_1\,\tau_2\r]_{O(r)}\,
\l(M_F\r)_{O(2)} = & \l[\tau\r]\; \pm \l(\frac{\tau}{2}\r),
\pm \l(\frac{\tau}{2} -1\r), \ldots, 0\;\;\mbox{or}\;\;\pm\frac{1}{2} \nn8
& & \oplus \l[\tau-1,1\r]\;\pm \l(\frac{\tau}{2}-1\r),
\pm \l(\frac{\tau}{2} -2\r), \ldots, 0\;\;\mbox{or}\;\;\pm\frac{1}{2} \nn8
& & \oplus \l[\tau-2,2\r]\;\pm \l(\frac{\tau}{2}-2\r),
\pm \l(\frac{\tau}{2} -3\r), \ldots, 0\;\;\mbox{or}\;\;\pm\frac{1}{2} \nn8
& & \oplus \ldots \nn8
& & \oplus \l[\tau-2\r]\;\pm \l(\frac{\tau}{2}\r),
\pm \l(\frac{\tau}{2} -1\r), \ldots \nn8
& & \oplus \ldots \;\;\;\;\;\;\;\;\;\;\;\;\;\;\;\;\;\;\;\;\;\;\;\;\;\;\;\;
\;\;\;\;\;\;\;\;\;\;\;\;\;\;\;(A6)
\earr
$$
\newpage
{\bf Table 1.} $\l[\tau\r]_{O(2r)} \rightarrow \l[\tau_1\,\tau_2\r]_{O(r)}\,
\l(M_F\r)_{O(2)}$ irrep reductions for $\tau \leq 6$. The results in the
table are verified using the dimension formulas \cite{Wy-70} for $r=5,6$,
$d(\l[\tau_1 \tau_2\r]_{O(5)}) = (2\tau_1+3)(2\tau_2+1)(\tau_1-\tau_2+1)
(\tau_1+\tau_2+2)/6$,
$d(\l[\tau_1 \tau_2\r]_{O(6)}) = (\tau_1+2)^2(\tau_2+1)^2(\tau_1-\tau_2+1)
(\tau_1+\tau_2+3)/12$. Used also is the formula $d(\l[\tau\r]_{O(2r)})=
{\footnotesize{\l(\barr{cc} \tau +2r-1 \\ \tau \earr\r)}} -
{\footnotesize{\l(\barr{cc} \tau +2r-3 \\ \tau-2 \earr\r)}}$ for any $r$.
\begin{center}
\begin{tabular}{rrr|rrr}
\hline
\\
$\l[\tau\r]_{O(2r)}$ & $\l[\tau_1\,\tau_2\r]_{O(r)}$ & $\l(M_F\r)_{O(2)}$
& $\l[\tau\r]_{O(2r)}$ & $\l[\tau_1\,\tau_2\r]_{O(r)}$ & $\l(M_F\r)_{O(2)}$
\nn8
\\
\hline
\\
$\l[0\r]$ & $\l[0\r]$ & 0 & $\l[5\r]$ & $\l[5\r]$ & $\pm\frac{5}{2}$,
$\pm\frac{3}{2}$, $\pm\frac{1}{2}$ \nn8
$\l[1\r]$ & $\l[1\r]$ & $\pm\frac{1}{2}$ & & $\l[41\r]$ &
$\pm\frac{3}{2}$, $\pm\frac{1}{2}$ \nn8
$\l[2\r]$ & $\l[2\r]$ & $\pm 1$, 0  & & $\l[32\r]$ &
$\pm\frac{1}{2}$ \nn8
& $\l[11\r]$ & $0$ & & $\l[3\r]$ &
$\pm\frac{5}{2}$, $\pm\frac{3}{2}$, $\l(\pm\frac{1}{2}\r)^2$ \nn8
& $\l[0\r]$ & $\pm 1$ & & $\l[21\r]$ &
$\pm\frac{3}{2}$, $\pm\frac{1}{2}$ \nn8
$\l[3\r]$ & $\l[3\r]$ & $\pm \frac{3}{2}$, $\pm \frac{1}{2}$ & & 
$\l[1\r]$ & $\pm\frac{5}{2}$, $\pm\frac{3}{2}$, $\pm\frac{1}{2}$ \nn8
& $\l[21\r]$ & $\pm \frac{1}{2}$ & $\l[6\r]$ & $\l[6\r]$ &
$\pm 3$, $\pm 2$, $\pm 1$, 0 \nn8
& $\l[1\r]$ & $\pm\frac{3}{2}$, $\pm \frac{1}{2}$ & & $\l[51\r]$ &
$\pm 2$, $\pm 1$, 0 \nn8
$\l[4\r]$ & $\l[4\r]$ & $\pm 2$, $\pm 1$, 0 & & $\l[42\r]$ &
$\pm 1$, 0 \nn8
& $\l[31\r]$ & $\pm 1$, 0 & & $\l[33\r]$ & 0 \nn8
& $\l[22\r]$ & 0 & & $\l[4\r]$ & $\pm 3$, $\pm 2$, $(\pm 1)^2$, $(0)^2$ \nn8
& $\l[2\r]$ & $\pm 2$, $\pm 1$, $(0)^2$ & & $\l[31\r]$ & $\pm 2$, $\pm 1$,
$(0)^2$ \nn8
& $\l[11\r]$ & $\pm 1$ & & $\l[22\r]$ & $\pm 1$ \nn8
& $\l[0\r]$ & $\pm 2$, 0 & & $\l[2\r]$ & $\pm 3$, $\pm 2$, $(\pm 1)^2$, 
0 \nn8
& & & & $\l[11\r]$ & $\pm 2$, 0 \nn8
& & & & $\l[0\r]$ & $\pm 3$, $\pm 1$ \nn8
\\
\hline
\end{tabular}
\end{center}

\newpage
\baselineskip=22pt
\begin{center}
{\bf Figure Caption}
\end{center}

Fig. 1. Typical energy spectrum in the $O(12) \supset O(6) \otimes O(2)$
limit of pnIBM for $N=6$ bosons with $M_F=-1$ (i.e. $N_\pi=2$, $N_\nu=4$).
The parameters in the energy formula (20) are chosen to be $a_1=30$ keV,
$a_2=-125$ keV, $a_3=35$ keV and $a_4=10$ keV. The $O(5)$ quantum numbers
$\l[v_1 v_2\r]$ are shown to the left of each energy level and to the right
shown are $L^\pi$ values. Energy levels for $v=6$ with 
$\l[\sigma_1 \sigma_2\r]=\l[6\r]$ and $\l[51\r]$ and $v=4$ with 
$\l[\sigma_1 \sigma_2\r]=\l[4\r]$ are shown in the figure.

\ed
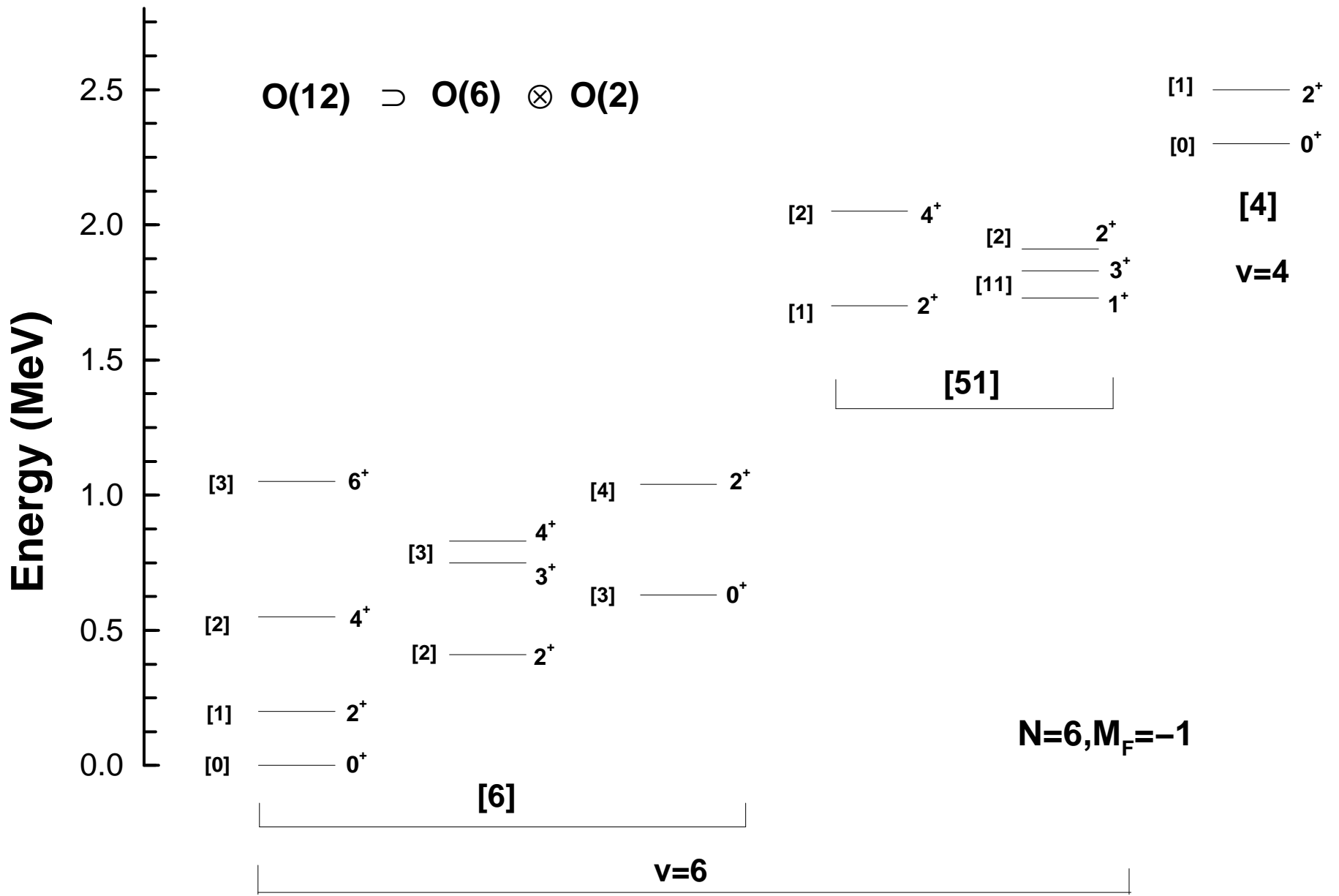
